\documentclass[seceqn]{elsart5p}

\usepackage{graphicx}
 
\graphicspath{{.}{./EPS/}}

\usepackage{amssymb}

\journal{Solid State Communications}

\newcommand{\ket}[1]{\left | \, #1 \right \rangle}

\begin{document}

\begin{frontmatter}



\title{Rashba interferometers: Spin-dependent single and
two-electron interference}


\author[IFS,MacD]{U. Z\"ulicke\corauthref{corinfo}}, 
\ead{u.zuelicke@massey.ac.nz}
\author[IFS]{A.~I. Signal}

\corauth[corinfo]{Corresponding author. {\it Tel.:\/} 
+64 6 350 5799 extn 7259, {\it fax:}\/ +64 6 350 5682.}

\address[IFS]{Institute of Fundamental Sciences, Massey University,
Private Bag 11~222, Palmerston North, New Zealand}
\address[MacD]{MacDiarmid Institute for Advanced Materials and
Nanotechnology, Massey University, Private Bag 11~222, Palmerston North,
New Zealand}

\begin{abstract}

Quantum transport in semiconductor nanostructures can be described theoretically in terms
of the propagation and scattering of electron probability waves. Within this approach, elements
of a phase-coherent electric circuit play a role similar to quantum-optical devices that can be
characterised by scattering matrices. Electronic analogues of well-know optical interferometers
have been fabricated and used to study special features of charge carriers in solids. We
present results from our theoretical investigation into the interplay between spin
precession and quantum interference in an electronic Mach-Zehnder interferometer with
spin-orbit coupling of the Rashba type. Intriguing spin-dependent transport effects occur,
which can be the basis for novel spintronic devices such as a magnet-less
spin-controlled field-effect transistor and a variety of single-qubit gates. Their 
functionality arises entirely from spin-dependent interference of each single input
electron with itself. We have also studied two-electron interference effects for the spin-dependent
Mach-Zehnder interferometer, obtaining analytical expressions for its two-fermion-state
scattering matrix. Using this result, we consider ways to generate two-electron output
states for which the Rashba spin-subband quantum number and the output-arm index
are entangled. Combining spin-dependent interference in our proposed Mach-Zehnder
interferometer with a projective charge measurement at the output enables entanglement 
generation. As our particular scheme involves tuneable spin precession, electric-field control
of entanglement production can be achieved.
\end{abstract}

\end{frontmatter}

\section{Introduction}
\label{introSec}

The quantum superposition principle governing the microscopic world can be most 
clearly demonstrated by interference effects involving quantum particles. A
famous~\cite{doubleSlit} example is Feynman's double-slit \textit{Gedankenexperiment\/}
with  electrons, which was later realised~\cite{tono:ajp:89} using an
electron-microscopy setup. While quantum interference and phase coherence have
long been a central issue in quantum optics~\cite{loudon}, mesoscopic electron
transport~\cite{mesotrans} has only comparatively recently provided a versatile
testing ground for these concepts in a solid-state context. Quantum phase coherence
of electrons in conductors results in well-studied bulk effects such as weak localisation 
and universal conductance fluctuations~\cite{imrybook}. However, the advent of
micro- and nanofabrication techniques has lifted the study of phase-coherent electron
transport onto a new level. It has made it possible, e.g., to observe
Aharonov-Bohm~\cite{ahabohm} oscillations of the electrical conductance through
mesoscopic rings~\cite{aboscmetal,aboscsemi} and enabled the design of the first
solid-state electron interferometers~\cite{bartinterfere,tuneabosc,elmABeffect}. By now,
electronic versions~\cite{amir:prl:94,basel:sci:99,liu:sci:99,electronMZ,NewelectronMZ}
of optical interferometers have been fabricated and studied in great detail. Interference
of electron waves scattered by impurities or gates in a semiconductor heterostructure
was imaged using scanning-probe microscopy~\cite{west:nat:01,ania:prl:05}. In
addition to providing us with a better fundamental understanding of quantum effects
in solids, the study of phase-coherent transport can also lead to the design of novel
electronic devices~\cite{newDattaBook}. In particular, linear-optics-inspired
architectures~\cite{milb:nat:01} for quantum information processing may eventually
enable the creation of solid-state flying qubits~\cite{been:prl:04}.

The expectation that greater functionality of electronic devices can be achieved
by utilising a charge carrier's spin degree of freedom is generating a lot of interest
in studying spin-dependent transport effects~\cite{sciencerev,lossbook,zut:rmp:04}.
Spin dependence can simply be introduced by the presence of magnetic materials.
Based in this concept, a spin-dependent electronic Fabry-P\'erot interferometer was
realised in a magnetic multilayer~\cite{spininterfere1,spininterfere2}. More recently,
state-of-the-art semiconductor engineering has made it possible to investigate the
interplay between phase coherence and spin dependence of electronic transport in
\emph{nonmagnetic\/}
nanostructures~\cite{shay:prl:02,kato:apl:05,mole:prl:06,koga:prb:06}.
Theoretical proposals for spin-dependent electronic
Aharonov-Bohm~\cite{nitta:apl:99,irene:prb-rc:03,kis:jap:03,peet:prb:04,frust:prb:04,bran:prb:04,peet:prb:06,chingray},
Sagnac~\cite{koga:prb:04}, Mach-Zehnder~\cite{uz:apl:04}, and Young
(double-slit)~\cite{uz:jsup:05} interferometers are based on
gate-tuneable~\cite{nitta:prl:97,schaep:prb-rc:97} spin precession~\cite{spinfet} due
to the Rashba effect~\cite{byra:jpc:84} that arises from structural inversion asymmetry
(SIA)~\cite{wink:prb:00} in semiconductor heterostructures. Spin-interference effects
due to Rashba spin splitting are present in quantum networks~\cite{gov:prl:04} and multiple
coherent scattering of electrons from intrinsic and artificial impurities~\cite{walls:prb:06}.

Here we provide a review and more details of our recent studies of single~\cite{uz:apl:04}
and two-particle~\cite{uz:apl:05} interference in spin-dependent Mach-Zehnder interferometers.
The following Section is devoted to a brief introduction to SIA-induced (Rashba) spin splitting,
discussing its salient features and relevance for spin-dependent interferometry. Single-particle
interference effects and their possible use in spintronics and quantum information processing
are discussed in Section~\ref{sec:SingInt}. We provide results for two-particle interference and
entanglement generation in Sec.~\ref{sec:DoubInt}. Our conclusions and an outlook to
future work are presented in Sec.~\ref{sec:Conc}. A detailed analytical expression for the
scattering matrix of the spin-dependent Mach-Zehnder interferometer discussed here is
given in the Appendix.

\section{Spin precession arising from SIA spin splitting}
\label{sec:Rashba}

\begin{figure}[t]
\begin{center}
\includegraphics*[width=1.6in]{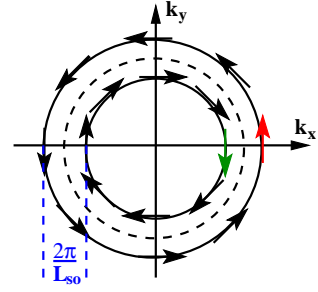} 
\end{center}
\caption{Fermi surface of a 2D electron system with SIA spin splitting. For any
propagation direction, two eigenstates exist with opposite spin and wave numbers differing by
$2\pi/L_{\mathrm{so}}$. The spin quantisation axis for eigenstates is always perpendicular to
their 2D wave vector and thus cannot be fixed globally for all eigenstates.}
\label{fig1}
\end{figure}

Two-dimensional (2D) electrons in asymmetric semiconductor heterostructures are subject to
a coupling of their spin and orbital degrees of freedom~\cite{rolandbook} which gives rise
to a correction
\begin{equation}\label{SIAso}
H_{\mathrm{R}} = \frac{\pi \hbar}{m L_{\mathrm{so}}} \, \left( \sigma_x p_y - \sigma_y p_x \right)
\end{equation}
to their kinetic-energy term in the Hamiltonian. Here $\sigma_x,\sigma_y$ denote Pauli matrices,
$p_x,p_y$ are the electron's in-plane momentum components, and the spin-precession
length~\cite{spinfet} $L_{\mathrm{so}}$ measures the strength of Rashba spin
splitting~\cite{byra:jpc:84} (and thus SIA of the heterostructure) in terms of a length scale.
Typical values are $L_{\mathrm{so}}\sim 400$~nm in InGaAs/InP~\cite{schaep:prb-rc:97}
and $\lesssim 100$~nm in HgTe/HgCdTe~\cite{buh:prb:04} heterostructures.
Eigenstates of the 2D free-electron Hamitonian including the SIA spin-orbit term~(\ref{SIAso})
are plane waves characterised by wave number $\vec k=(k_x,k_y)$. They are also eigenstates
(with eigenvalue $\sigma=\pm 1$) of spin projection in the in-plane direction that is
perpendicular to $\vec k$. The energy dispersion, given by
\begin{equation}
E_\sigma(\vec k) = \frac{\hbar^2}{2 m} \left( k - \sigma \, \frac{\pi}{L_{\mathrm{so}}} \right)^2
- \frac{\pi^2\hbar^2}{2 m L_{\mathrm{so}}^2} \quad ,
\end{equation}
is spin-split but, unlike in the case of Zeeman splitting, SIA spin splitting does not result in a net
magnetisation of the 2D system. At a fixed energy such as the Fermi energy, electron states are
grouped into two circles in $\vec k$-space as shown in Fig.~\ref{fig1}. For each propagation
direction, two states exist that have opposite spin polarisation perpendicular to $\vec k$ and
wave vector differing by $2\pi/L_{\mathrm{so}}$.

\begin{figure}[b]
\begin{center}
\includegraphics*[width=1.6in]{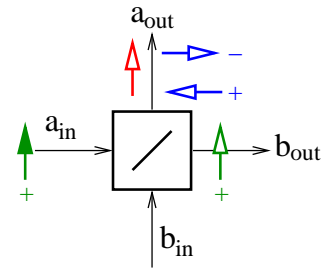} 
\end{center}
\caption{Scattering of SIA-spin-split states at a beam splitter. The electron wave incoming
from the left and its transmitted partial wave do not experience spin precession because
they are in the spin-split eigenstate with $\sigma=+1$. This is not true for the reflected
partial wave. As the latter's spin is parallel to its propagation direction, the reflected wave is a
superposition of $\sigma=\pm 1$ eigenstates and its spin will precess. Within the scattering
theory of quantum transport that we employ, the beam splitter is a four-terminal device
characterised by a scattering matrix that relates quantum spinor amplitudes $a_{\mathrm{in}},
b_{\mathrm{in}}$ at the input terminals to those ($a_{\mathrm{out}}, b_{\mathrm{out}}$) at
the output.}
\label{fig2}
\end{figure}
There are two relevant consequences of SIA spin splitting for electron-wave interference in
nanostructures. Firstly, the Fermi-wave-vector difference for opposite-spin states moving
in the same direction results in different dynamical phases acquired by the corresponding
two partial waves. As a result, a superposition of eigenstates injected in a wave guide will
evolve during propagation to yield a precession of spin~\cite{crook:prl:05,kato:apl:05b} with
spatial period $L_{\mathrm{so}}$. Gate-tuneability of SIA that gives rise to the spin splitting
can be used to adjust~\cite{spinfet} the spin-precession length and thus manipulate the spin
state of electrons emerging from the wave guide. Secondly, the dependence of eigenstates'
spin quantisation axis on the propagation direction implies that spin-conserving scattering of
partial waves can induce spin precession of incident SIA-spin-split eigenstates. Figure~\ref{fig2}
illustrates this fact for a beam splitter. No spin precession occurs for the electron state incident
from the left and its transmitted partial wave emitted to the right because their spin is polarised
perpendicular to the propagation direction. In contrast, the reflected partial wave has the same
spin state as the incoming electron but moves in a direction that is perpendicular to the line of
incidence. With spin and propagation direction now parallel for the reflected partial wave, it is a
superposition of SIA eigenstates for this outgoing terminal and spin precession commences.

\section{Spin-dependent Mach-Zehnder interferometer}
\label{sec:SingInt}

\begin{figure}[t]
\begin{center}
\includegraphics*[width=1.7in]{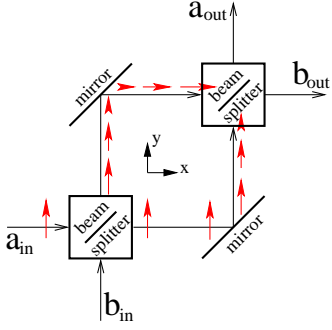} 
\end{center}
\caption{Spin-dependent electronic Mach-Zehnder interferometer. Beam separation and spin
precession for an electron wave incident from the left in an SIA-spin-split eigenstate are
indicated. Our theoretical description of this device is based on its scattering matrix that can
be obtained by suitable combination of the spin-resolved scattering matrices for its constituent
parts.}
\label{fig3}
\end{figure}
The Mach-Zehnder (MZ) interferometer is routinely used in quantum optics~\cite{loudon}. Its
basic structure, consisting of two beam splitters and two mirrors, is illustrated in Fig.~\ref{fig3}.
In the following, we consider electronic realisations of the interferometer, where quantum
point contacts could be used for beam splitters~\cite{basel:sci:99,liu:sci:99,electronMZ} and
mirror reflection occurs from etched walls. The presence of SIA spin splitting introduces
spin precession of electron waves in the interferometer, as indicated in Fig.~\ref{fig3}.
Back-gate voltages could be used~\cite{nitta:prl:97,schaep:prb-rc:97} to adjust the
spin-precession length $L_{\mathrm{so}}$ \emph{uniformly} in the interferometer structure. Note
that, in contrast to other theoretical
proposals~\cite{irene:prb-rc:03,kis:jap:03,egues:prl:02,angik:apl:06}, we do not employ
local variations in the SIA spin splitting for our device concepts.

The function of each linear-optical device, as well as the acquisition of spin-dependent
dynamical phases for travelling electron waves, can be described mathematically by appropriate
scattering matrices~\cite{rashbascatt} that relate outgoing and incoming quantum probability-wave
amplitudes. We assume straight wave guides to be attached to each terminal of a beam
splitter/mirror, making it possible to choose SIA-spin-split eigenstates as our scattering
states. Using standard notation suitably generalised to our situation with spin-dependent
scattering (see Ref.~\cite{uz:apl:04}, especially Fig.~1 and Eq.~(2), for our conventions), the
${\mathcal S}$-matrices for an ideal symmetric beam splitter, ideal mirror(s), and
one-dimensional interferometer wave guides are, respectively,
\begin{eqnarray}
{\mathcal S}_{\mathrm{bs}} &=& \left(
\begin{array}{cccc}
 \frac{i}{2} & -\frac{1}{2} & \frac{1}{\sqrt{2}} & 0 \\
 -\frac{1}{2} & \frac{i}{2} & 0 & \frac{1}{\sqrt{2}} \\
 \frac{1}{\sqrt{2}} & 0 & \frac{i}{2} & \frac{1}{2} \\
 0 & \frac{1}{\sqrt{2}} & \frac{1}{2} & \frac{i}{2}
\end{array}
\right) \quad , \\
{\mathcal S}_{\mathrm{m}} &=& \left(
\begin{array}{cccc}
 \frac{i}{\sqrt{2}} & -\frac{1}{\sqrt{2}} & 0 & 0 \\
 -\frac{1}{\sqrt{2}} & \frac{i}{\sqrt{2}} & 0 & 0 \\
 0 & 0 & \frac{i}{\sqrt{2}} & \frac{1}{\sqrt{2}} \\
 0 & 0 & \frac{1}{\sqrt{2}} & \frac{i}{\sqrt{2}}
\end{array}
\right) \quad , \\
{\mathcal S}_{(x,y)} &=& \left(
\begin{array}{cccc}
 0 & 0 & e^{-\frac{i \pi  y}{L_{\mathrm{so}}}} & 0 \\
 0 & 0 & 0 & e^{\frac{i \pi  y}{L_{\mathrm{so}}}} \\
 e^{-\frac{i \pi  x}{L_{\mathrm{so}}}} & 0 & 0 & 0 \\
 0 & e^{\frac{i \pi  x}{L_{\mathrm{so}}}} & 0 & 0
\end{array}
\right) \quad .
\end{eqnarray}
Due to the special MZ-interferometer design, the outputs from the first beam splitter become
orthogonal mirror inputs. Essentially, each partial-wave beam can be considered to hit on a
different side of a two-faced mirror before interfering with the other at the second beam splitter.
We can therefore use ${\mathcal S}_{\mathrm m}$ to simultaneously represent the effect of both
mirrors. The scattering matrix of an entire MZ-interferometer structure is then obtained by simple
multiplication of appropriate ${\mathrm S}$-matrices:
\begin{equation}
{\mathcal S}_{\mathrm{MZ}} = {\mathcal S}_{\mathrm{bs}} \, 
{\mathcal S}_{\mathrm{sw}} \, {\mathcal S}_{(w, h)} \, {\mathcal S}_{\mathrm{sw}} \,
{\mathcal S}_{\mathrm{m}} \, {\mathcal S}_{\mathrm{sw}} \, {\mathcal S}_{(w, h)} \,
{\mathcal S}_{\mathrm{sw}} \, {\mathcal S}_{\mathrm{bs}} \, .
\end{equation}
We omitted an unimportant phase factor $\exp\{i k_E (w + h)\}$ where $k_E$
is the average of the two Fermi wave vectors for SIA-spin-split bands. The matrix
${\mathcal S}_{\mathrm{sw}}=\sigma_x \otimes \sigma_x$ converts quantum amplitudes at
the outgoing terminals of any interferometer part into incoming ones for the next. $w$ and
$h$ are the spatial extensions of the MZ interferometer in $x$ and $y$ directions, respectively.
The general analytic expression for ${\mathcal S}_{\mathrm{MZ}}$ is given in the Appendix.
Here we focus on a several special cases that are of potential interest to
spintronics~\cite{lossbook} and quantum information processing~\cite{qcqip}.

(i)~$w$ and $h$ are integer multiples of the spin-precession length $L_{\mathrm{so}}$. In this
case, the MZ interferometer transmits any incoming state perfectly, except for a possible
overall sign change of quantum amplitudes. This is the default situation~\cite{MZgeneral} for a
rectangularly shaped conventional Mach-Zehnder interferometer where the optical path lengths
in the two interferometer arms are equal. Phase shifts due to reflections at beam splitters/mirrors
then result in constructive (desctructive) interference in the transmitting (reflecting) channel.
In the following, we demonstrate deviations from this canonical behaviour, which are entirely
due to spin dependence of electron-wave interference.

(ii)~$w$ is an integer but $h$ is a half-integer multiple of $L_{\mathrm{so}}$. To be specific,
let us assume $w=m \, L_{\mathrm{so}}$ and $h=(n+1/2)L_{\mathrm{so}}$. Then 
${\mathcal S}_{\mathrm{MZ}}$ reduces to
$$ (-1)^{\left(m+n+\frac{1}{2}\right)} \left(
\begin{array}{cccc}
 0 & 0 & 1 & 0 \\
 0 & 0 & 0 & -1 \\
 0 & i & 0 & 0 \\
 -i & 0 & 0 & 0
\end{array}
\right) \quad . $$
The interferometer is still totally transmitting but, in the horizontal ($x$) direction, transmission
of each SIA-split eigenstate is associated with a spin flip. Note that the interferometer's
extension in this direction is an integer multiple of $L_{\mathrm{so}}$; thus no spin flip
could have been achieved for any spin state by transmission through a wave guide of the
same length. More fundamentally, however, this spin flip occurs for SIA-split eigenstates (whose
spin is polarised in the vertical direction). Such states would not precess and, hence, no spin flip
could ever be engineered for these by transmission through a horizontal wave guide. On the
other hand, the state with spin polarised in horizontal direction (i.e., parallel to the propagation
direction) would be faithfully transmitted by this interferometer configuration. Obviously, an
analogous situation where the behaviour of horizontal and vertical transmission of spin-split
eigenstates is exchanged occurs when $w$ is a half-integer and $h$ an integer multiple of the
spin-precession length.

(iii)~Both $w$ and $h$ are half-integer multiples of $L_{\mathrm{so}}$. The MZ-interferometer
scattering matrix for this case can be written, upto an overall sign, as ${\mathcal S}_{\mathrm
m}^\ast$, i.e., it is the complex conjugate of the mirror's ${\mathrm S}$-matrix. Evidently, the
interferometer is now purely reflecting, which constitutes a drastic departure from the canonical
behaviour of a MZ interferometer~\cite{MZgeneral}. Interestingly, cases (i) and (iii) can be
realised for any quadratic interferometer (i.e., having $w=h$) by suitable adjustment of
$L_{\mathrm{so}}$. This enables the design of a magnet-less spin-controlled field-effect
transistor~\cite{uz:apl:04,uz:jsup:05}.

(iv)~$w$ is an integer multiple and $h$ a quarter-integer multiple of $L_{\mathrm{so}}$. Upto
an overall sign, the MZ interferometer's ${\mathrm S}$-matrix for this case can be written as
$$
\left(
\begin{array}{cccc}
 0 & 0 & (-1)^{-1/4} & 0 \\
 0 & 0 & 0 & (-1)^{1/4} \\
 \frac{1}{\sqrt{2}} & \frac{1}{\sqrt{2}} & 0 & 0 \\
 -\frac{1}{\sqrt{2}} & \frac{1}{\sqrt{2}} & 0 & 0
\end{array}
\right) \quad .
$$
Thus the interferometer is fully transmitting and realises a (pseudo-)Hadamard
transformation~\cite{qcqip} in the horizontal channel. An analogous configuration with a
Hadamard transformation for the vertical channel can be obtained by having $w$ instead of $h$
be a quarter-integer multiple of $L_{\mathrm{so}}$.

(v)~Any serial combination of two (or more) MZ interferometers can be discussed as well.
Geometric considerations suggest that additional mirrors would have to be inserted, and
spin-dependent phase shifts be accounted for. We choose to neglect the latter (possible in
the limit when the path length between the two interferometers is much smaller than the
spin-precession length) but include the mirrors.  As an example, we give results for an
interferometer of type~(iv) in series with one of type~(iii). Again neglecting a possible overall
sign change of quantum amplitudes, the scattering matrix for the two-interferometer setup is
$$
i \left(
\begin{array}{cccc}
 0 & 0 & 0 & (-1)^{3/4} \\
 0 & 0 &  (-1)^{-3/4} & 0 \\
 \frac{1}{\sqrt{2}} & \frac{1}{\sqrt{2}} & 0 & 0 \\
 \frac{1}{\sqrt{2}} & -\frac{1}{\sqrt{2}} & 0 & 0
\end{array}
\right) \quad .
$$
Here the horizontally transmitting channel is subject to a Hadamard transformation.
Scattering states in the vertical channel experience a spin flip. In contrast, spin eigenstates for 
projection parallel to the propagation direction in the vertical channel are eigenstates of the
transmission matrix.

For the present discussion, it shall suffice to present the above examples. Inspection of the
general scattering matrix for spin-dependent MZ interferometers, which is given in the Appendix,
can be used to design yet more devices.

\section{Spin-dependent two-particle interference and entanglement generation}
\label{sec:DoubInt}

In the previous Section, we discussed phenomena arising from electron probability waves
interfering with themselves in a MZ-interferometer setup. The question immediately arises
whether the interplay between spin-dependent interference and Fermi statistics can give rise to
nontrivial two-particle effects. After all, it is well-known that the Pauli principle affects multi-particle
scattering events, e.g., at a beam splitter~\cite{loud:pra:98}. We investigated this question by
means of a two-particle scattering approach. Its basic ingredient is the definition of a suitable
basis of two-particle scattering states. The quantum numbers to use for distinguishing
single-particle states are the interferometer terminal (a or b) and the SIA-spin-split subband
quantum number ($\sigma=\pm 1$). Thus the single-particle Hilbert space is four-dimensional.
A straightforward consideration yields that the corresponding two-particle Hilbert space for
fermionic particles is six-dimensional. We found it useful to employ a basis of Bell states similar
to the one introduced in Ref.~\cite{eck:ann:02},
\begin{eqnarray}
\ket{\chi_1}&=&\frac{1}{\sqrt{2}}\left(c^\dagger_{\mathrm{a}+}
c^\dagger_{\mathrm{a}-} + c^\dagger_{\mathrm{b}+} 
c^\dagger_{\mathrm{b}-} \right) \ket{0} \quad , \\ 
\ket{\chi_2}&=&\frac{1}{\sqrt{2}}\left(c^\dagger_{\mathrm{a}+} 
c^\dagger_{\mathrm{b}+} - c^\dagger_{\mathrm{a}-} 
c^\dagger_{\mathrm{b}-} \right)  \ket{0} \quad , \\
\ket{\chi_3}&=&\frac{1}{\sqrt{2}}\left(c^\dagger_{\mathrm{a}+} 
c^\dagger_{\mathrm{b}-} + c^\dagger_{\mathrm{a}-} 
c^\dagger_{\mathrm{b}+} \right)  \ket{0} \quad , \\
\ket{\chi_4}&=&\frac{i}{\sqrt{2}}\left(c^\dagger_{\mathrm{a}+} 
c^\dagger_{\mathrm{a}-} - c^\dagger_{\mathrm{b}+}
c^\dagger_{\mathrm{b}-} \right)  \ket{0} \quad , \\
\ket{\chi_5}&=&\frac{i}{\sqrt{2}}\left(c^\dagger_{\mathrm{a}+} 
c^\dagger_{\mathrm{b}+} + c^\dagger_{\mathrm{a}-} 
c^\dagger_{\mathrm{b}-} \right)  \ket{0} \quad , \\
\ket{\chi_6}&=&\frac{i}{\sqrt{2}}\left(c^\dagger_{\mathrm{a}+} 
c^\dagger_{\mathrm{b}-} - c^\dagger_{\mathrm{a}-} 
c^\dagger_{\mathrm{b}+} \right)  \ket{0} \quad ,
\end{eqnarray}
where the operator $c^\dagger_{\alpha\sigma}$ creates an electron at the
Fermi energy in the interferometer arm $\alpha$ and with SIA spin-split subband
quantum number $\sigma$. Any two--electron state $\ket{\mathrm{i}} $
($\ket{\mathrm{o}}$) at the input (output) can be expressed as a linear combination
of these basis states. Using our previous results for spin-resolved single-particle
scattering matrices at beam splitters, mirrors, etc.\, and straightforwardly implementing
the Pauli principle, we derived~\cite{uz:apl:05} the $6\times 6$ scattering matrices that
can be used to calculate the two-particle state generated at the output of a linear-optical
device from any incoming state that is expressed as a linear superposition of the $\ket{\chi_j}$.
Suitable combination of the individual devices' two-particle scattering matrices, in a way that is
similar to the single-particle case, yields the corresponding two-particle scattering matrix for
the entire interferometer. Its complicated analytical expression is omitted here.

\begin{figure}[b]
\begin{center}
\includegraphics*[width=2.3in]{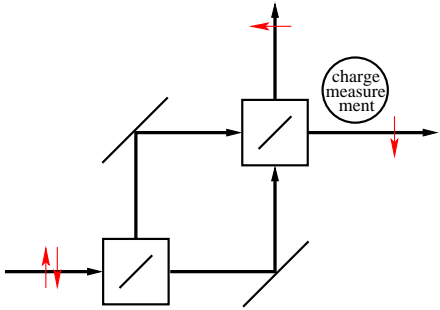} 
\end{center}
\caption{Two-particle scattering and entanglement generation at the spin-dependent
Mach-Zehnder interferometer. A non-entangled input state incident from the left is
transformed into an output state that has a finite probability for single-particle occupancy
in each output arm. A charge measurement at the device's output enables us to select
such cases non-deterministically. It turns out that the corresponding output states are
always maximally entangled in the SIA-spin-split subband quantum number.}
\label{fig4}
\end{figure}
\begin{figure}[t]
\begin{center}
\includegraphics*[width=2.3in]{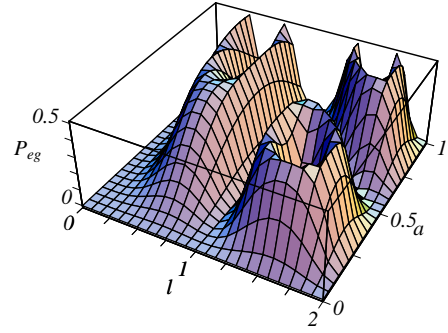} 
\end{center}
\caption{Efficiency $P_{\mathrm{eg}}$ of entanglement generation by the procedure illustrated in
Fig.~\ref{fig4}, calculated for rectangular MZ interferometers of width $l L_{\mathrm{so}}$ and
aspect ratio $a$.}
\label{fig5}
\end{figure}

Again, a variety of special cases can be considered that are distinguished by different
shapes/sizes of the interferometer, as well as different two-particle input configurations.
In the following, we focus on the situation where a two-particle product state with
double-occupancy is incident in one interferometer arm, while the other input terminal is
kept floating. See Figure~\ref{fig4} for an illustration. To be specific, we consider
the state $\ket{\mathrm{i}_{\mathrm{dbl}}} = c^\dagger_{\mathrm{a}+}
 c^\dagger_{\mathrm{a}-}\ket{0}$.
The output state generated from it by the MZ interferometer will be a
superposition of states having single electron occupancy in each output arm
with others that have a finite amplitude for double occupancy in one of them.
Let us assume that we have installed a device to measure the electron number (i.e.,
charge) in one of the output terminals~\cite{been:prl:04}. Performing this measurement each
time we send in two electrons with opposite spin at the input channel, we will obtain 
values 0, 1, or 2 with certain frequencies. The charge measurement leaves the spin and
orbital wave functions of output states unaffected, but it allows us to filter out those states
where exactly one electron is scattered into each of the output arms. Calculation of the concurrence~\cite{woot:prl:98} proves that, by preparing the input state
$c^\dagger_{\mathrm{a}+} c^\dagger_{\mathrm{a}-} \ket{0}$ and choosing states at the
interferometer output for which exactly one electron lives in each output arm, we obtain
two-electron states that are maximally entangled in the SIA-spin-split subband
quantum number $\sigma$ and the output-terminal index (a and b). The efficiency
$P_{\mathrm{eg}}$ for this nondeterministic scheme of entanglement generation is given
by the probability of producing the output state with single-electron occupancy per output
arm from the input state $ \ket{\mathrm{i}_{\mathrm{dbl}}}$. This quantity turns out to depend
on the interferometer size and geometry. As in the previous Section, we consider a rectangular MZ
interferometer of width $w$ and height $h$. Figure~\ref{fig5} shows a plot of $P_{\mathrm{eg}}$ for
a certain range of dimensionless parameters $l=w/L_{\mathrm{so}}$ and $a=h/w$. The largest
possible value of $P_{\mathrm{eg}}$ is 50\%, i.e., the same as the efficiency of entanglement
generation at a single beam splitter~\cite{bos:prl:02}. Unlike such a beam splitter, however, the MZ
interferometer can be tuned by an external gate voltage, which controls the parameter $l$ by
adjusting $L_{\mathrm{so}}$~\cite{nitta:prl:97,schaep:prb-rc:97}, between maximum and zero
entanglement generation. This realises a switcheable entangler. 

\section{Conclusions and outlook}
\label{sec:Conc}

We presented theoretical results for spin-dependent single and two-particle interference in an
electronic Mach-Zehnder interferometer with a uniform electric-field-tuneable spin splitting.
Analytic expressions are obtained for scattering matrices, enabling us to study realisations of
spintronic devices and quantum gates. With single electrons as input, spin-dependent
interference enables the design of a magnet-less spin-controlled field-effect switch. Single-qubit
gates such as the Hadamard gate can also be realised. Introduction of an additional charge measurement
at the interferometer output makes it possible to generate entangled mobile-electron spins
nondeterministically.  The efficiency of entanglement generation can be tuned between
0 and 50\% by adjusting the spin-precession length using external gate voltages. Future work
will be focused on applying methods presented here to study other promising spin-dependent
interference devices, e.g., modeled after the Michelson and Franson~\cite{franson}
interferometers.

\section*{Acknowledgements}
UZ is supported by the Marsden Fund Council from Government funding,
administered by the Royal Society of NZ.

\appendix

\onecolumn
\section{Analytical expression for the general MZ-interferometer scattering matrix}
\label{detailMZ}

\begin{equation}
{\mathcal S}_{\mathrm{MZ}} = \left(
\begin{array}{cccc}
 -\frac{i \sin \left(\frac{\pi h }{L_{\mathrm{so}}}\right) \sin \left(\frac{\pi  w}{L_{\mathrm{so}}}
 \right)}{\sqrt{2}} &
   -\frac{\sin \left(\frac{\pi h}{L_{\mathrm{so}}}\right) \sin \left(\frac{\pi  w}{L_{\mathrm{so}}}\right)}{\sqrt{2}} &
   -e^{-\frac{i \pi h}{L_{\mathrm{so}}}} \cos \left(\frac{\pi  w}{L_{\mathrm{so}}}\right) & \cos \left(\frac{\pi h
   }{L_{\mathrm{so}}}\right) \sin \left(\frac{\pi  w}{L_{\mathrm{so}}}\right) \\
 -\frac{\sin \left(\frac{\pi h}{L_{\mathrm{so}}}\right) \sin \left(\frac{\pi  w}{L_{\mathrm{so}}}\right)}{\sqrt{2}} &
   -\frac{i \sin \left(\frac{\pi h}{L_{\mathrm{so}}}\right) \sin \left(\frac{\pi  w}{L_{\mathrm{so}}}\right)}{\sqrt{2}} &
   -\cos \left(\frac{\pi h}{L_{\mathrm{so}}}\right) \sin \left(\frac{\pi  w}{L_{\mathrm{so}}}\right) & -e^{\frac{i \pi h
   }{L_{\mathrm{so}}}} \cos \left(\frac{\pi  w}{L_{\mathrm{so}}}\right) \\
 -e^{-\frac{i \pi  w}{L_{\mathrm{so}}}} \cos \left(\frac{\pi h}{L_{\mathrm{so}}}\right) & -\cos \left(\frac{\pi 
   w}{L_{\mathrm{so}}}\right) \sin \left(\frac{\pi h}{L_{\mathrm{so}}}\right) & -\frac{i \sin \left(\frac{\pi
 h  }{L_{\mathrm{so}}}\right) \sin \left(\frac{\pi  w}{L_{\mathrm{so}}}\right)}{\sqrt{2}} & \frac{\sin \left(\frac{\pi h
   }{L_{\mathrm{so}}}\right) \sin \left(\frac{\pi  w}{L_{\mathrm{so}}}\right)}{\sqrt{2}} \\
 \cos \left(\frac{\pi  w}{L_{\mathrm{so}}}\right) \sin \left(\frac{\pi h}{L_{\mathrm{so}}}\right) & -e^{\frac{i \pi 
   w}{L_{\mathrm{so}}}} \cos \left(\frac{\pi h}{L_{\mathrm{so}}}\right) & \frac{\sin \left(\frac{\pi h }{L_{\mathrm{so}}}\right) \sin
   \left(\frac{\pi  w}{L_{\mathrm{so}}}\right)}{\sqrt{2}} & -\frac{i \sin \left(\frac{\pi h}{L_{\mathrm{so}}}\right) \sin
   \left(\frac{\pi  w}{L_{\mathrm{so}}}\right)}{\sqrt{2}}
\end{array}
\right)
\end{equation}








\end{document}